\documentclass{emulateapj}
\usepackage{amsmath,amssymb}
\usepackage{graphicx,mathptm} 
\usepackage{times} 
\usepackage{natbib}
\bibpunct{[}{]}{,}{n}{}{;}
\usepackage{multirow}

\newcommand{\be}{\begin{equation}}
\newcommand{\ee}{\end{equation}}
\newcommand{\bea}{\begin{eqnarray}}
\newcommand{\eea}{\end{eqnarray}}

\shortauthors{Yan, Lazarian, \& Schlickeiser}
\begin{document}    
\title{Cosmic Ray streaming from SNRs and gamma ray emission from nearby molecular clouds}
\author{Huirong Yan\altaffilmark{1}, A. Lazarian\altaffilmark{2}, R. Schlickeiser\altaffilmark{3}}
\altaffiltext{1}{Kavli Institute of Astronomy \& Astrophysics, Peking University, Beijing, 100871, China, hryan@pku.edu.cn}
\altaffiltext{2}{Department of Astronomy, University of Wisconsin, Madison, WI 53706, USA}
\altaffiltext{3}{ Institute for Theoretical Physics IV, Ruhr University,  Bochum, 44780, Germany}

\begin{abstract}
High-energy gamma ray emission has been detected recently from supernovae remnants (SNRs) and their surroundings. The existence of molecular clouds near some of the SNRs suggests that the gamma rays originate predominantly from p-p interactions with cosmic rays accelerated at a closeby SNR shock wave. Here we investigate the acceleration of cosmic rays and the gamma ray production in the cloud  self-consistently by taking into account the interactions of the streaming instability and the background turbulence both at the shock front and in the ensuing propagation to the clouds. We focus on the later evolution of SNRs, when the conventional treatment of the streaming instability is valid but the magnetic field is enhanced due to either Bell's current instability and/or due to the dynamo generation of magnetic field in the precursor region. We calculate the time dependence of the maximum energy of the accelerated particles. This result is then used to determine the diffusive flux of the runaway particles escaping the shock region, from which we obtain the gamma spectrum consistent with observations. Finally, we check the self-consistency of our results by comparing the required level of diffusion with the level of the streaming instability attainable in the presence of turbulence damping. The energy range of cosmic rays subject to the streaming instability is able to produce the observed energy spectrum of gamma rays.     
\end{abstract}


\section{introduction}

Cosmic rays below the knee are generally believed to originate from shock acceleration at  the supernova remnants (SNRs). Although the process was proposed decades ago, it has not been  easy to constrain the model from the observationally. Recent advances on gamma ray observations  provide us a possibility to study the acceleration and ensuing propagation processes in detail. In particular,  the molecular clouds in the vicinity of a SNR are good testbeds because of their high density. Indeed, clouds illuminated by cosmic rays accelerated at a nearby SNR can be bright gamma ray sources (Aharonian \& Atoyan 1996; Fatuzzo et al. 2006; Gabici et al. 2009).   

Gamma rays are produced through the decay of neutral pions generated in p-p collisions (hadronic origin) or through inverse Compton interactions by energetic electrons with ambient photon target fields such as the cosmic microwave background radiation (leptonic origin). It is still under debate whether the observed gamma ray emissions are of the hadronic or leptonic origin although the hadronic scenario is favored by many authors because of the evidence of strongly amplified magnetic fields near SNRs (see the review by Blasi 2010 and references therein).

We consider the gamma ray emission from massive molecular clouds near SNRs, motivated, in particular, by the recently detected gamma ray emission from the SNR W28. As an old SNR, the emission from W28 is perceived of purely hadronic origin since the life time of high energy electrons is short. The observed gamma emission from W28 indicates that the flux of cosmic rays is much enhanced there compared to the typical Galactic values. This places a constraint on the cosmic ray diffusion. The enhancement of scattering corresponding to a reduction ($1-10\%$) of the CR diffusion coefficient compared to Galactic mean has been suggested by a few earlier models to match the observations (Fujita et al. 2009, Ohira et al. 2011, Li \& Chen 2010). No physical justification on how and why the scattering is boosted has been provided, however. In addition, earlier studies made a few serious assumptions including the Bohm diffusion and a phenomenological power law evolution of the maximum energy accelerated at a certain epoch. We shall reinvestigate this problem by incorporating  a proper physical description of the relevant processes, i.e.  the shock acceleration in the presence of the streaming instability and nonlinear damping processes by background turbulence. The slower diffusion implies enhanced wave perturbations, which may arise from streaming instability (see, e.g., Longair 2002). Earlier work has shown that streaming instability is limited by background turbulence (Yan \& Lazarian 2002; Farmer \& Goldreich 2004; Beresnyak \& Lazarian 2008). This could be the determinative factor for the maximum energy attainable at the shock front  as suggested by Ptuskin \& Zirakashvili (2005). However, the isotropic Kolmogorov scaling for the turbulence, adopted by Ptuskin \& Zirakashvili (2005), is not applicable to MHD turbulence. Moreover, if the enhanced scattering in the vicinity of SNRs as indicated by the observations are due to the increased flux of cosmic rays there, we need to examine whether the flux of accelerated particles is sufficient to induce high enough growth rates of the streaming instability to overcome the nonlinear damping by background turbulence.

In this paper, we apply our present day understanding of the interaction between the streaming instability and the background turbulence to the modeling of the gamma ray emission from molecular clouds near SNRs. We shall treat the problem in a self-consistent way by comparing the streaming level that is allowed by the preexisting turbulence and the required diffusion for the CRs. In \S 2 we determine the maximum energy of accelerated particles at shocks, in \S3 we calculate the spectrum of cosmic rays in the vicinity of SNRs, in \S4 we obtain the spectrum of gamma ray produced by the p-p interactions of the escaping cosmic rays in a nearby cloud and compare it with observations, \S5, we check the self-consistency of our result by comparing the required the diffusion near SNRs and the streaming level allowed in the presence of turbulence. The discussion and summary are provided in \S6 and \S 7. 
In the Appendix we provide a list of notations used in the paper. 

\section{Maximum energy of CRs accelerated at the shocks }

Diffusive shock acceleration of energetic CR particles relies on the crucial process of amplification of MHD turbulence so that particles can be trapped at the shock front long enough to be accelerated to the high energy observed. One of the most popular scenarios that has been adopted in the literature is the streaming instability generated by the accelerated particles. However, in the highly nonlinear regime the fluctuations of magnetic field arising from the streaming
instability get large and the classical treatment of the streaming instability is not applicable. We circumvent the
problem by proposing that the field amplification we consider does not arise from the streaming
instability, but is achieved earlier through other processes, e.g. the interaction of the shock precursor with density perturbations preexisting in the interstellar medium (Beresnyak, Jones, \& Lazarian 2009). Due to the
resonant nature of the streaming instability, the perturbations $\delta B$ arising from it are more efficient
in scattering CRs compared to the large scale fluctuations produced by non-resonant mechanisms, e.g.
the one in Beresnyak et al. (2009). Therefore in this paper, we limit our discussions to the regime of $\delta B \la B_0$, where $B_0$ is the magnetic field that has already been amplified in the precursor region\footnote{The effective $B_0$ is therefore renormalized and can be much larger than the typical field in ISM (see, e.g., Diamond \& Malkov 2007).}.  

When particles reach the maximum energy at a certain time, they escape and the growth of the streaming instability stops. Therefore we can obtain the maximum energy by considering the stationary state of the evolution. The steady state energy density of the turbulence $W(k)$ at the shock is determined by

\begin{equation}
(U\pm v_A)\nabla W(k) = 2 (\Gamma_{cr}-\Gamma_d)W(k),
\label{wave}
\end{equation}
where $U$ is the shock speed, and the term on the l.h.s. represents the advection of turbulence by the shock flow. $v_A\equiv B_0/\sqrt{4\pi nm}$ and $n$ are the Alfven speed and the ionized gas number density of the precursor region, respectively. The plus sign represents the forward propagating Alfven waves and the minus sign refers to the backward propagating Alfven waves. The terms on the r.h.s. describes the wave amplification by the streaming instability and damping with $\Gamma_{cr}, \Gamma_d$ as the corresponding growth and damping rates of the wave.  The distribution of accelerated particles at strong shocks is $f(p)\propto p^{-4}$. If taking into account the modification of the shock structure by the accelerated particles, the CR spectrum becomes harder. Assume the distribution of CRs at the shock is $f_0(p)\propto p^{-4+a}$. The nonlinear growth was studied by Ptuskin \& Zirakashvili (2005). 

It was demonstrated by Schlickeiser \& Shalchi (2008) that waves can grow or damp at the shock precursor depending on the the spatial boundary conditions. If assuming an equal amount of forward and backward waves at the shock front, the forward wave will be growing and the backward wave will be efficiently damped in the upstream region. We therefore neglect the backward moving wave mode and consider here only the growing forward moving mode, which is the one that effectively contributes to the particle scattering. The generalized growth rate of streaming instability then is

\begin{eqnarray}
\Gamma_{cr}&=&\frac{12\pi^2 q^2v_A\sqrt{1+A^2}}{c^2k}\nonumber\\
&\times& \int^\infty_{p_{res}} dp p\left[1-\left(\frac{p_{res}}{p}\right)^2\right]D\left|\frac{\partial f}{\partial x}\right|, 
\label{general}
\end{eqnarray}
where $q$ is the charge of the particle, c is the light speed, $p_{res}=ZeB_0\sqrt{1+A^2}/c/k_{res}$ is the momentum of particles that resonate with the waves.  $A=\delta B/B_0$ is wave amplitude normalized by the mean magnetic field strength $B_0$.
\begin{equation}
D=\sqrt{1+A^2}v r_g/3/A^2(>k_{res})
\label{crdiff}
\end{equation}
is the diffusion coefficient of CRs, $v$ and $r_g$ are the velocity and Larmor radius of the CRs. The distribution function of CRs is

\begin{eqnarray}
f_0&=&\frac{3\xi nm U^2H(p_{max}-p)}{4\pi c\phi(p_{max}) (mc)^ap^{4-a}}, \nonumber\\
\phi(p_{max}) &=& \int_0^{p_{max}/mc} \frac{dy y^a}{\sqrt{1+y^2}},
\label{f0}
\end{eqnarray}
where $\xi$ measures the ratio of CR pressure at the shock and the upstream  momentum flux entering the shock front, $m$ is the proton rest mass, and $p_{max}$ is the maximum momentum accelerated at the shock front. $H(p)$ is the Heaviside step function.
 
In the planar shock approximation, the distribution of accelerated particles is

\begin{equation}
f_1(p,x) =f_0(p) \exp\left(-(U+v_A) \int_0^x \frac{dx_1}{D(p,x_1)}\right)
\label{f1}
\end{equation}
at the upstream of the shock ($x\geq0$) and $f=f_0$ at the downstream. Insert Eqs.(\ref{crdiff}, \ref{f0},\ref{f1}) into  Eq.(\ref{general}), one gets the following growth rate of the upstream forward moving wave at x=0,
\begin{equation}
\Gamma_{cr}(k)=\frac{C_{cr}\xi U^2(U+v_A)k^{1-a}}{(1+A^2)^{(1-a)/2}cv_A\phi(p_{max})r_0^a} 
\label{growth}
\end{equation}
where $C_{cr}=9/2/(4-a)/(2-a)$, $r_0=m c^2/q/B_0$.
The linear damping is negligible since the medium should be highly ionized. In fully ionized gas, there is nonlinear Landau damping, which, however, is suppressed due to the reduction of particles' mean free path in the turbulent medium (see Yan \& Lararian 2011; Brunetti \& Lazarian 2011); we therefore neglect this process here. Background turbulence itself can cause nonlinear damping to the waves (Yan \& Lazarian 2002). Unlike hydrodynamical turbulence, MHD turbulence is anisotropic with eddies elongated along the magnetic field. The anisotropy increases with the decrease of the scale (Goldreich \& Sridhar 1995). At the scales of the Larmor radii $r_L$ of the cosmic rays, which are also the wavelengths $1/k$ of the waves induced by the streaming instability, the scale disparity becomes very large with $k_\bot \gg k^t_\|$,  where we use $k^t_\|$ to distinguish the parallel component of the turbulence wave packet wavenumber $k^t_\|$ from the parallel wavenumber of the growing wave $k_\|$. In MHD turbulence, wave packet cascades after it travels a distance of $1/k^t_\| \sim L^{1/3}k_\bot^{-2/3} \gg 1/k_\bot$. On the other hand, the instability grows fastest for the most parallel wave ($k\sim k_\|$) allowed in a turbulence medium with their wave numbers satisfying $k_\bot/k_\| \sim \delta B/B\sim (k_\bot L)^{-1/3}$. Because of the scale disparity, $k_\| > k_\bot \gg k^t_\|$, the nonlinear damping rate in MHD turbulence is less than the wave frequency $k_\| v_A$, and it is given by (Farmer \& Goldreich 2004; Yan \& Lazarian 2004; Beresnyak \& Lazarian 2008)

\begin{equation}
\Gamma_d \sim \sqrt{k/L} v_A,
\label{damping}
\end{equation}
where L is the injection scale of background turbulence, and the $k$ is set by the resonance condition $k \sim k_\| \sim 1/r_L$.
Inserting Eqs.(\ref{growth},\ref{damping}) into Eq.(\ref{wave}) and adopting $U\partial W/\partial x \approx U^2 W/D$ in the case of efficient wave  amplification, one gets

\begin{equation}
\frac{3U^2A^2}{2v(1+A^2)}+v_A/\sqrt{kL} =\frac{C_{cr}\xi U^2(U+v_A)}{cv_A\phi(p_{max})(kr_0)^a(1+A^2)^{(1-a)/2}}.
\label{wave2}
\end{equation}

There are various models for the diffusive shock acceleration. We consider here the escape-limited acceleration. In this model, particles are confined in the region near the shock where turbulence is generated. Once they propagate far upstream at a distance $l$ from the shock front, where the self-generated turbulence by CRs fades away, the particles escape and the acceleration ceases.  The characteristic length that particles penetrate into the upstream is $D(p)/U$. The maximum  momentum is reached when $D(p)/U\simeq l/4$\footnote{The factor 1/4 arises from the following reason. As pointed out by Ostrowski \& Schlickeiser (1996), the spectrum is steepened for small l, i.e., $l U/D(p) \la 4$}. Assuming $l\propto R_{sh}$, then the maximum momentum of particles accelerated during the Sedov phase is determined by the condition

\begin{equation}
D(p_{max})=\kappa UR_{sh} ,
\label{Dshock}
\end{equation}
where $\kappa<1$ is a numerical factor, see table 2. From equations (\ref{crdiff},\ref{Dshock}), we get
\begin{equation}
\frac{p_{max}}{mc} = \frac{3\kappa A^2 U R_{sh}}{\sqrt{1+A^2}v r_0}
\label{gmax}
\end{equation}

Insert Eq.(\ref{gmax}) into Eq.(\ref{wave2}), we get for $A<1$\footnote{We neglect a factor of $\sqrt{1+A^2}$ here.}

\begin{eqnarray}\frac{p_{max}}{mc}&=&\left[\left(-v_A\sqrt{\frac{1}{r_0 L}}+\sqrt{\frac{v_A^2}{r_0 L}+\frac{2C_{cr}a\xi U^3(U+v_A)}{\kappa r_0R_{sh}cv_A}}\right) \left(\frac{\kappa R_{sh}}{U}\right)\right]^2,\nonumber\\
A&=&\frac{p_{max}r_0}{\sqrt{18}\kappa mU R_{sh}}\sqrt{1+\sqrt{1+36 \left(\frac{\kappa mU R_{sh}}{p_{max}r_0}\right)^2}},
\label{gmax_general}
\end{eqnarray}
where $\phi(p_{max})$ is approximated by $(p_{max}/mc)^a/a$.

\begin{figure}
\includegraphics[width=0.95\columnwidth,  height=0.24\textheight]{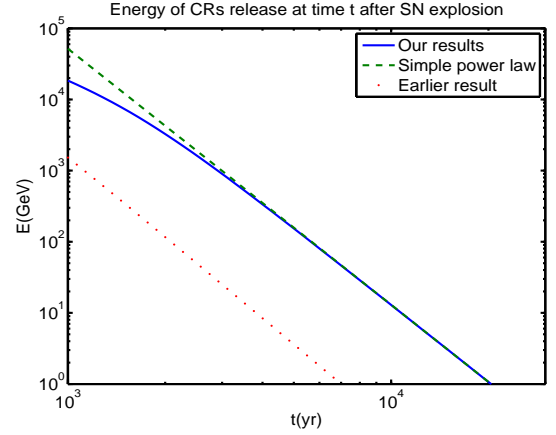}
 \caption{The energy of CRs that are released at the shock at time t in the Sedov phase. Our result shows that the often assumed power law solution (see Gabici et al. 2009; Ohira et al. 2010) is only realized in asymptotic regime as described in Eqs.(\ref{lowshockU},\ref{gmax_solution}). It is also larger than the earlier result (dotted line) in Ptuskin \& Zirakashvili (2005) where the damping of the waves by background turbulence is overestimated.}
\label{Emax} 
\end{figure}

In the limit of low shock velocity, 
\begin{eqnarray}
v_A\ll U\ll  c\left[\left(\frac{v_A}{c}\right)^3\frac{\kappa R_{sh}}{2 L C_{cr}a\xi}\right]^{1/4},
\label{lowshockU}
\end{eqnarray}
we get 
\begin{eqnarray}
\frac{p_{max}}{mc}&=&(C_{cr}\xi U^3)^2\frac{a^2 L}{r_0c^2v_A^4}.
\label{gmax_solution}
\end{eqnarray}
At the Sedov phase ($t>t_{sed}\equiv 250(E_{51}/(n_0U_9^5))^{1/3}$yr), where $E_{51}=E_{SN}/10^{51}$erg and $U_9=U_i/10^9$cm/s are the total energy of explosion and the initial shock velocity, the evolution of shock radius and speed are governed by

\begin{eqnarray}R_{sh}&=&4.3(\epsilon_{51}/n_0)^{1/5} t_{kyr}^{2/5}{\rm pc}, \nonumber\\
U&=&1.7 \times 10^3 (\epsilon_{51}/n_0)^{1/5} t_{kyr}^{-3/5} {\rm km/s}.
\label{Sedov}
\end{eqnarray}
From Eqs.(\ref{gmax_solution},\ref{Sedov}), we see that $p_{max} \propto t^{-\delta}$, and $\delta=18/5$. In Fig.\ref{Emax}, we plot $p_{max}/(mc)$ vs. the time t since supernova explosion. The solid line represents the results from Eqs.(\ref{gmax_general}). As we see, at earlier epoch when advection and streaming instability are both important, the evolution of $p_{max}$ does not follow a power law. For comparison, we also put a power law evolution in the same figure as depicted by Eq.(\ref{gmax_solution})  (dashed line). 
Our result is also larger than that obtained by Ptuskin \& Zirakashvili (2005) since the wave dissipation rate is overestimated in their treatment.

\section{The cosmic ray distribution near SNRs}

At a given time, only particles with $p>p_{max}$ can escape the shock front. Adopting the simplified approximation, 

\begin{equation}
D(p)\begin{cases}\ll R_{sh} U, & p<p_{max}\\ 
\gg  R_{sh} U, & p>p_{max}
\end{cases}
\end{equation}
at a given energy $E = cp$, there is a one to one correspondence between the CR momentum and the CR escape time $t_{esp}$ (or radius $R_{esp}=(1+\kappa)R_{sh}$) in spite of the fact that the acceleration is continuous during the shock expansion since $p_{max}(t)$ reaches $p$. If the maximum momentum has a power law dependence on t, one can easily gets $t_{esp}\propto p^{-1/\delta}$.

Since we consider the later stage of SNR acceleration, the shock radius cannot be neglected and the point source assumption is not applicable in general.  One needs to take into account the spatial distribution of the sources. 
Nevertheless, in the case that the diffusion distance  $R_d=2\sqrt{D(t_{age}-t_{esp})}$ is larger than the radius $R_{esp}$, the distribution function of energetic proton at a distance $r$ can be described as if the CRs were from a point source  (see, e.g., Ohira et al. 2011)

\begin{equation}
F(E) \approx \frac{f(E)}{\pi^{3/2}R_{d}^3} \exp\left[-\left(\frac{r}{R_{d}}\right)^2\right],
\label{dist_finl}
\end{equation}
The diffusion coeffcient $D=\chi D_{ISM}$, where $D_{ISM}=10^{28}\sqrt{E/10{\rm GeV}}{\rm cm}^2$/s (Gabici et al. 2009; Berezinskii et al. 1990).  $f(E)$ is the spectrum of runaway particles that is integrated over shock expansion.

We adopt the routine from Ohira et al. (2010) here. If the accelerated particles at a given time is
\begin{equation}
f(p,t)dpdt =K(t) \left(\frac{p}{m c}\right)^{-2+a} \exp\left[-\frac{p}{p_{max}(t)}\right]d \ln t dp,
\end{equation}
then the general spectrum of protons dispersed from the accelerator is 
\begin{equation}
f(p) \propto \frac{p^{-1+a} K(p_{max}^{-1}(p))}{p^{-1}_{max}(p)[dp_{max}/dt]_{t=p^{-1}_{max}(p)}}
\label{spectrum_ave}
\end{equation}
where $p^{-1}_{max}(p)$ is the inverse function of $p_{max}(t)$. $K(t)$ is a normalization factor and can be estimated from
\begin{equation}
P_{CR}\approx \frac{4\pi}{3}\int_{m_pc}^{p_{max}(t)}dp v p^3 f_0(p) =\xi nm U^2.
\end{equation}

In the case of flatter spectrum at the shock front, i.e., $a>0$, $K(t)\propto U^2R_{sh}^3p^{-a}$. Evaluation of  $dp_{max}/{dt}$ from Eq.(\ref{gmax_general}) shows that it can be well represented by a power law in spite of the fact that $p_{max}$ does not follow an exact power law. Insert $K(t)$ and $dp_{max}/{dt}\simeq p_{max}/t$ into Eq.\ref{spectrum_ave}, we get a universal power spectrum $f(E) \propto \eta E_{SN}E^{-s}$ with $s=2$ regardless of the  flatter original spectrum at the shock front\footnote{The influence of CR cooling on the spectrum index is negligible (Ohira et al. 2010).}. $\eta$ is the fraction of SN energy converted into CRs. We obtain a similar cosmic ray momentum spectrum as earlier authors (e.g., Ptuskin \& Zirakashvili 2005; Ohira et al. 2010) without assuming the power law evolution of $p_{max}(t)$. $p_{max}(t)$ in our paper was obtained by considering the balance of the wave growth from the streaming instability, the advection by the shock flow as well as wave damping by the background turbulence. The latter is due to the nonlinear interaction between the wave and the turbulence, and has not been properly accounted for. Ptuskin \& Zirakashvili (2005) considered a similar process, but assumed isotropy of the MHD turbulence which is not valid. 

\begin{figure}
\includegraphics[width=0.95\columnwidth,  height=0.24\textheight]{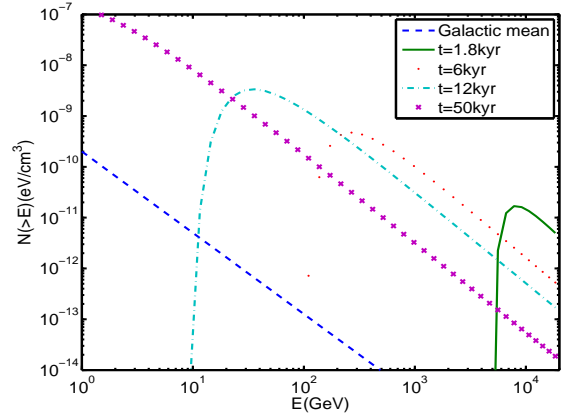}
 \caption{The spectrum of CRs at a distance $r=12$pc after 1800 (solid line), 6000 (dotted line), 12000 (dashdot line), 50000 years (cross line). The Galactic mean is plotted as a reference (dashed line).}
 \label{CRspectra}
\end{figure}

Although time evolution of $p_{max}(t)$ does not follow a power law at high energies \footnote{It is a good approximation for $E_{max}\la 1000$GeV (or $t > 4$kyr) in the case we consider (see Fig.\ref{Emax}).}, the final CR spectrum $F(E)$ at a later age of Sedov phase and a distance $r> R_d$ is well reproduced by the power law form if the diffusion coefficient $D(p)$ has a power law dependence on particle momentum p. This is because high energy CRs are produced at earlier epoch $t_{esp}\ll t_{age}$ so that $R_d$ can be well approximated by $R_d\simeq 2\sqrt{Dt_{age}}$. Adopting the set of parameters for W28 as listed in Tables 1,2, we get the flux of CRs $E^2F(E)$. In Fig.\ref{CRspectra}, we plot the flux of CRs at different epochs after the SN explosion. All the values of the parameters except that for turbulence are compatible with those from earlier work ( Fujita et al. 2009; Gabici et al. 2010; Li \& Chen 2010). We adopt a turbulence injection scale of $30pc$, which is consistent with the scenario that interstellar turbulence is injected from supernovae explosions. The CR energy at each peak corresponds to the maximum momentum $p_{max}(t)$ reached at each epoch t. The galactic background is also plotted for comparison. Apparently, the CR intensities are dramatically enhanced in the vicinity of SNRs as compared to the Galactic mean intensities. 

\begin{table*}
\caption{Physical parameters of W28}
\addtolength{\tabcolsep}{-3pt}
\begin{tabular}{ccccccccccc}
\hline
\hline
r(kpc)&$E_{SN}(10^{51})$erg&$M_c (M_\odot$)&$t_{age}$(kyr)&$U_i$(km/s)&L(pc)&n(cm$^{-3}$)&$R_c(pc)$&T(K)&$B_{0}(\mu$G)&$B_{cav}(\mu$G)\\
\hline
1.8&0.5&$4\times 10^4$&50&5500&30&8&12&$10^6$&200&2\\
\hline
\hline
\end{tabular}
\end{table*}

\section{gamma ray flux in the vicinity of SNRs}

\begin{figure}
\includegraphics[width=1\columnwidth,
  height=0.24\textheight]{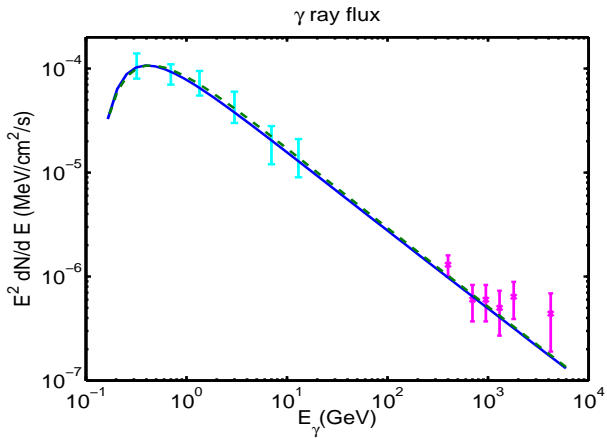}
 \caption{The spectrum of Gamma ray emission from W28. The Fermi data are shown as dotted points (Abdo et al. 2010), and the H.E.S.S. data are plotted as 'x' points (Aharonian et al. 2008) with error bars. Solid line is our result.}
 \label{gamma}
\end{figure}

In the case $F(E) $ follows a power law distribution, the pion gamma ray emissivity is given by (see Aharonian \& Atoyan 1996)

\begin{equation}
q(E_\gamma) \approx \frac{16\pi f_\pi^{s-1}}{s^2}\sigma_{pp}F(E_\gamma)c\eta_A ,
\label{pp}
\end{equation}
where $E_\gamma \approx E_p/10$ is the corresponding $\gamma$ ray energy, $\sigma_{pp}\approx 30\times \left[0.95+0.06\ln (E_p/{GeV})\right]$mb is the cross section for pp collisions at $E_p$, $f_\pi$ is the fraction of energy that is transferred from parent protons to secondary pions, $\eta_A\simeq 1.4-1.5$ is a parameter to account for the contribution from both cosmic rays and the interstellar gas (Dermer 1986). The total flux then is

\begin{equation}
\frac{dN_\gamma}{dE_\gamma}=\frac{M_c q(E)}{4\pi d^2 m}, 
\label{gm_flux}
\end{equation}
where $M_c,\,d$ are the mass and the distance of the cloud. 
Combining Eqs.(\ref{dist_finl},\ref{pp},\ref{gm_flux}), we obtain the flux of gamma ray emission as shown in Fig.\ref{gamma}, where our result is plotted against both Fermi and H.E.S.S. data. The GeV and TeV data are adopted from Abdo et al. (2010) and Aharonian et al. (2008), respectively. Our result produces a power law spectral index $\sim 2.75$, showing that the steepening of the spectrum of particle can be naturally explained by the propagation effect. Indeed, similar fits have been also obtained by other  models, e.g., Ohira et al (2011), Li \& Chen (2010). We did not need, however, either to assume Bohm diffusion and phenomenological power law evolution of the momentum of the escaping partciels as in Ohira et al. (2011), or a steeper spectrum for the escaping particles as in Li \& Chen (2010).  A decrease of the spatial diffusion coefficient by a factor of $\chi=0.05$ comparable to earlier works (Gabici 2010, Li \& Chen 2010) is inferred here, which can originate from streaming instability. Estimates will be provided below to demonstrate that the instability can grow for the energy range we consider. 

\section{Enhanced scattering and streaming instability near SNRs}

Our results show that the local scattering of CRs has to be enhanced by an order of magnitude $\chi =0.05$ in order to produce the amount of $\gamma$ ray emission observed. A natural way to increase the scattering rate is through the streaming instability. We provide here a self-consistency check by examining whether the streaming instability operates in the presence of nonlinear damping by the background turbulence. We envisage a local low density, high temperature cavity with $B_{cav}=2\mu$G surrounding the supernova subjected to strong UV radiation and stellar wind (see Fugita et al. 2009), so that there is no ion-neutral damping in this case. Only nonlinear damping exists (see eq.\ref{damping}). The growth rate in the linear regime is
\footnote{We neglect the nonlinear Landau damping, which is suppressed in turbulence due to decrease of mean free path.}. 

\begin{equation}
\Gamma_{gr}=\Omega_0\frac{N(\geq E)}{n}\left(\frac{v_s}{v_A}-1\right),
\end{equation}
where $v_s$ is the streaming speed of CRs. The growth rate should overcome the damping rate (eq.\ref{damping}) for the instability to operate. The condition $\Gamma_{gr}>\Gamma_d$ leads to
\begin{equation}
v_s > v_A \left(1+\frac{n v_A}{N \Omega_0\sqrt{r_gL}}\right)
\end{equation}

\begin{figure}
\includegraphics[width=0.95\columnwidth,  height=0.24\textheight]{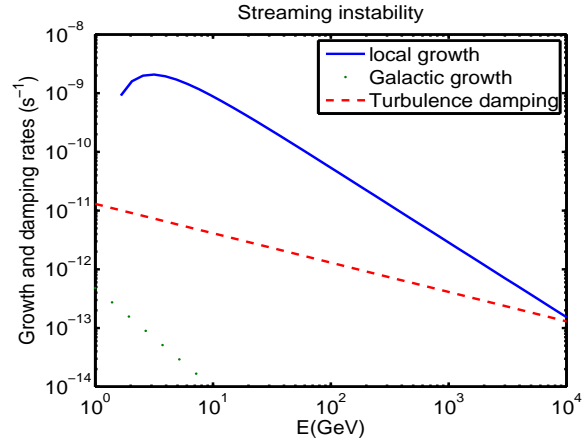}
 \caption{The growth and nonlinear damping rates of streaming instability. With the locally enhanced flux, the growth rate of streaming instability becomes much larger than the mean Galactic value so that it can overcome the nonlinear damping by turbulence for a wide energy range. This is consistent with our earlier treatment in which streaming instability plays an essential role in the cosmic ray diffusion near SNRs.}
 \label{rates}
\end{figure}

The spatial diffusion coefficient adopted here, $D \approx v_s L = \chi D_{ISM}$, satisfies  this requirement. The growth and damping rates are compared in Fig.\ref{rates}. We see that the streaming instability works in the energy range needed to produce the observed $\gamma$ ray emission, proving that our results are self-consistent.

Note that the case we consider here is different from the general interstellar medium discussed in Yan \& Lazarian (2004) and Farmer \& Goldreich (2004), namely, the local cosmic ray flux near SNRs is much enhanced (see Fig.2). Consequently, the growth rate of the streaming instability becomes high enough to overcome the damping rate by the preexisting turbulence in the considered 
energy range.

\begin{table}
\caption{Model parameters adopted in the paper}
\begin{tabular}{cccccc}
\hline
\hline
a&$\chi$&$\eta$&$ \kappa$&$\xi$&$\alpha$\\
\hline
0.1$\sim 0.3$&$\sim$0.05&$\sim 0.3$&$0.04\sim 0.1$& $0.2\sim$0.4& 0.5\\
\hline
\hline
\end{tabular}
\end{table}

\section{Discussion}

Gamma rays have been detected from molecular clouds near SNRs. They are believed to the result of interactions with the energetic particles accelerated at SNR shocks, and thus provide the best probe to both the shock acceleration acceleration and propagation processes of cosmic rays near SNRs. To determine the spectrum of CRs that generates the gamma ray emissions, it is necessary to  establish the time dependence of the energy of escaping particles. Earlier it has 
been done assuming that only CRs with $p>p_{max}(t)$ can escape and that the time dependence is a power law (see Gabici et al 2010, Ohira et al. 2011). In this paper, we derive our results on the basis of the well motivated and tested physical model of the streaming instability in the presence of background turbulence. With this model
we obtain $p_{max}(t)$ which is determined by the interaction of streaming instability at the upstream and the background turbulence. 
We limit the discussions to the later evolution of SNR acceleration which is stage of the observed gamma ray sources. In this case, the classical formula for the growth of streaming is valid.  

The spectral fit at high gamma ray energies with H.E.S.S. data is less optimal. High energy cosmic rays are generated earlier when strong magnetic amplification is needed. As we pointed out earlier, the classical treatment of streaming is invalid and moreover, the field amplification may well be due to other processes, eg., the current instability (Bell 2004) and/or turbulent density perturbations interacting with the shock precursor (Beresnyak et al.  2009).  

Another crucial ingredient  that determines the CR flux at the clouds is the spatial diffusion coefficient of CRs escaping from SNRs. It has been speculated recently that this spatial diffusion is suppressed due to CR induced instabilities to match the observed level of gamma ray fluxes (see Gabici et al 2010). No specific study has been provided so far. Here we place a constraint to the level of streaming by applying our present understanding of turbulence damping of waves (Yan \& Lazarian 2004, Farmer \& Goldreich 2004, Beresnyak \& Lazarian 2008). By balancing the growth rate of the streaming instability with the turbulence damping rate, we obtain not only the energy range at which the streaming instability operates, but also the lower limit of the CR spatial diffusion coefficient. These results are found to be consistent with the required enhanced scattering indicated from the interpretation of the gamma ray data.

For the spectral index of the the CR power law spectrum, we took into account the hardening of the accelerate particles at the shock front. This does not contradict the fact that the spectrum of runaway particles is steeper as it is a time-averaged result. For the same reason, we do not consider that the decrease of the compression ratio would contribute to the steepening of the power law spectrum of the accelertaed CRs.  In fact, as long as $s<2$, the spectral index of runaway particles becomes uniformly 2. The main steepening at a distance $r$ from the SNR is due to the propagation effect. In the case when $r\gtrsim R_d> R_{esp}$, the point source is a good approximation for SNR, and the spectrum is steepened by $p^{3\alpha/2}$ if $D\propto p^{\alpha}$. 

The diffusion coefficient $D$ near the shock is assumed to be of the form $D=\chi 10^{28}\sqrt{E/10{\rm GeV}}{\rm cm}^2$/s following earlier treatments (see Gabici et al 2010, Ohira et al. 2011).  Undoubtedly, solving the diffusion coefficient surrounding SNRs taking into account both streaming instability and background turbulence is an important step toward a complete self-consistent picture.  It will be one of our future endeavors. Another simplification we made concerns the magnetic field strength at the shock. For the later stage of Sedov phase evolution, we assumed an already increased magnetic field strength $\sim 100\mu$G, and did not consider the specifics of earlier amplification process, which is widely believed to be present. This process is important for earlier shock evolution and acceleration of ultra high energy CRs. It, however, is a subject of intensive debates and is definitely beyond the  scope of our current paper.

The set of physical parameters and model parameters we used are listed, respectively, in tables 1\&2. The physical condition we adopted for W28 is close to that used in literatures ( Fujita et al. 2009; Gabici et al. 2010; Li \& Chen 2010). As for the model parameters, we have tight constraints only for the diffusion coefficient ($\alpha, \chi$), similar to those obtained in earlier literatures (Gabici et al. 2010; Li \& Chen 2010) and that for the energy of the SN explosion $\eta$. The results are not so sensitive to the other model parameters, as shown in table 2.      

\section{Summary}

We investigated the acceleration of particles at the later stage of SNR evolution and the escape of these particles. We calculated the flux of the escaping CRs at a molecular cloud in the vicinity of SNRs. We quantitatively took into account the competition of 
turbulence generation by the streaming instability of CRs and background turbulence damping, both, near the shock waves and in the ensuing propagation of CRs from SNRs. The resulting gamma ray spectrum from the p-p interactions is compared with both Fermi and H.E.S.S. data. Our main results are: 

\begin{itemize}

\item The streaming instability plays a crucial role for cosmic ray acceleration, particularly at the later stage of SNRs. The competition of turbulence generation by the streaming instability and turbulence damping determines the maximum energy of the accelerated particles.

\item The spectrum of runaway particles follows a universal power law of $f(E) \propto E^{-2}$ if the original spectrum index $s$ at the shock front is $s\leq 2$ for the later Sedov phase. The main steepening at a distance from the SNR is due to the propagation effect.

\item The flux of CRs is increased by several orders of magnitude compared to the mean Galactic value, creating enough turbulence by the streaming instability in the vicinity of the shock which overcomes the damping by background turbulence for the energy range considered.

\item The enhanced scattering by the instability is enough to reproduce the observed gamma ray flux from massive molecular clouds near the SNR. 

\end{itemize}

\section*{Acknowledgments}
We acknowledge the constructive and helpful comments by the referee. HY is supported by 985 grant from Peking U and NSFC grant  AST -11073004. AL acknowledges the support by the NSF grant 0808118, Center for magnetic Self Organization. This work was completed during the stay of A.L. as
Alexander-von-Humboldt-Preistr\"ager at the Ruhr-University at Bochum. The work of RS is partially funded by the Deutsche
Forschungsgemeinschaft through grant Schl 201/23-1 and the
German Ministry for Education and Research
(BMBF) through Verbundforschung Astroteilchenphysik (grant 05A11PC1).

\appendix

\section{List of Notations}
The notation we used is listed in Table \ref{notations}.

\begin{table}
\label{notations}
\begin{tabular}{|c|r|}
\hline
A  & Normalized wave amplitude $\delta B/B_0$\\
a& hardening of the CR spectrum at the shock front\\
$B_0$ & mean magnetic field at the shock in the later Sedov phase\\
$B_{cav}$ &inercloud magnetic field strength\\
$\delta B$& wave amplitude\\
c & light speed\\
d& distance of the molecular cloud from observer\\
D& diffusion coefficient of CRs\\
E& CR energy\\
$E_{SN}$& supernova explosion\\
f& distribution function of CRs\\
$f_\pi$& fraction of energy transferred from parent protons to pions\\
k& wave number\\
K(t) & Normalization factor of CR distribution function\\
L& the injection scale of background turbulence\\
m& proton rest mass\\
$M_c$ & cloud mass\\ 
n& intercloud number density\\ 
$N_\gamma$ & $\gamma$ ray flux\\
p& CR's momentum\\
$p_{max}$& the maximum momentum accelerated at the shock front\\
$P_{CR}$ & CR pressure\\
q & charge of the particle\\
r& distance from SNR centre\\
$R_c$& the distance of the molecular cloud from the SNR centre\\
$r_g$ & Larmor radius of CRs\\
$R_d$ & diffusion distance of CRs\\
$R_{sh}$ & shock radius\\
$R_{esp},\,t_{esp}$& the escaping distance/time of CRs\\ 
s& 1D spectrum index of CR distribution\\
t& time since supernova explosion\\
$t_{age}$ & the age of SNR\\
$t_{sed}$ & the time at which SNR enters the Sedov phase\\
U& shock  speed\\
$U_i$ &initial shock velocity\\
v& particle speed\\
$v_s$ & streaming speed of CRs\\
W& wave energy\\
$\alpha$& power index of D with respect to particle momentum p\\
$\chi$&reduction factor of D with respect to $D_{ISM}$\\
$\delta$& power index of $p_{max}$ with respect to t\\
 $\eta$&fraction of SN energy converted into CRs\\
 $\eta_A$& a numerical factor in Eq.\ref{pp}\\
$\Gamma_{cr}$& the growth rate of streaming instability\\
$\Gamma_d$& wave damping rate\\
 $ \kappa$&ratio of diffusion length to shock radius\\
 $\Omega_0$ & the Larmor frequency of nonrelativistic protons\\
 $\sigma_{pp}$ &cross section for pp collision\\
 $\xi$& the ratio of CR pressure to fluid ram pressure\\
\hline
\end{tabular}
\end{table}

\end{document}